\documentclass{rjparticle}
\usepackage{graphicx}

\newcommand{\miktex}{\hbox{Mik\kern-.15em\TeX}}

\title{Duality condition for $s$- and $t$-channel exchange \\in nucleon-nucleon scattering}
\author[1,2]{M. I. Krivoruchenko}
\author[3]{Amand Faessler}
\affil[1]{
Institute for Theoretical and Experimental Physics$\mathrm{,}$ B. Cheremushkinskaya 25 \\ 
117218 Moscow, Russia }
\affil[2]{Department of Nano-$\mathrm{,}$ Bio-$\mathrm{,}$ Information and Cognitive Technologies\\ 
Moscow Institute of Physics and Technology$\mathrm{,}$ 9 Institutskii per. \\ 
141700 Dolgoprudny$\mathrm{,}$ Moscow Region$\mathrm{,}$ Russia}
\affil[3]{Institut f\"{u}r Theoretische Physik der Universit\"{a}t T\"{u}bingen$\mathrm{,}$ Auf der Morgenstelle 14 \\ 
D-72076 T\"{u}bingen, Germany}

\begin{document}
\maketitle
\begin{abstract}
We specify conditions under which the nucleon-nucleon interaction, 
based on the $t$-channel meson-exchange mechanism, is equivalent to an interaction 
generated via an $s$-channel exchange of six-quark bags.
The duality is possible provided the alternation of zeros and poles of the non-dispersive part of $D$ function takes place
in the normalization where the imaginary part of $D$ is non-negative 
and the CDD poles are the only poles of $D$. 
\end{abstract}

\section{Introduction}

The fact that elementary particles are associated with the poles of $S$ matrix was realized in the 1950s. 
Stable particles correspond to the poles for real values of energy and are 
identified with the asymptotic states of quantum field theory.
Most of the poles are located at complex energies. 
These states are interpreted as resonances. Resonances, if they are not too far from the unitary cut, 
are observed experimentally as peaks in the cross sections.
The most complete relationship between the singularities of $S$ matrix as an analytic function and the Feynman
diagrams was established by Landau \cite{LAND}.

Quark models predict the existence of exotic hadrons. 
The experimental searches for exotic mesons and baryons have not been successful. 
In the late1970s, Jaffe and Low \cite{JALO79} proposed 
to look for signs of exotic hadrons in $P$ matrix, but not $S$ matrix. 
$P$ matrix can be constructed from the known scattering phase and its properties can then be examined. 
In this sense, the $P$-matrix poles can be observed in about the same way as the $S$-matrix poles corresponding to the resonances, however, 
the difference being that $P$ matrix is a model-dependent object.
In any case, the understanding came out that the set of $S$-matrix poles is, perhaps, 
too limited to accommodate all of the (quasi-) particle states 
that play a dynamical role in the strong interaction. 
\footnote{On the other hand, the set of poles of $ S $-matrix is too broad: 
Among the isolated poles there is a subset of "spurious poles" that do not correspond 
to particles (see e.g. \cite{BAZZ69}).}
The poles of $P$ matrix specify stable particles, resonances, and the so-called "primitives". 
Jaffe and Low proposed to identify exotic hadrons with the primitives, i.e., $P$-matrix poles
that do not manifest themselves as poles of $S$ matrix.

A dynamic scheme of $P$ matrix as an extension of the Dyson model \cite{DYSO57}
was proposed by Simonov \cite{SIMO81}. 
Compound states of the model are associated with simple zeros of $D$ function \cite{MIKH10} 
\begin{equation}
D(s)=0.
\label{DZERO}
\end{equation}
A fraction of the zeros specifies bound states and resonances. 
The zeros on the unitary cut are associated with the primitives.
Primitives do not generate the singularities of $S$ matrix,
\begin{equation}
S(s)=\frac{D(s-i0)}{D(s+i0)},
\label{DSMAT}
\end{equation}
because the zeros occur in pairs in the denominator and the numerator and cancel each other. 
The effect is similar to the disappearance of the singularity 
from the amplitude when the resonance width tends to vanish.

In the scattering problem, there are discrete values of energy, that seem to play special role
in the dynamics,
not being singular points of $S$ as an analytic function. 
Those values identify states, the primitives, which are not asymptotic states and thus can not be 
registered by detectors straight out.
In an attempt to isolate them experimentally, 
it would be futile to look for peaks in the cross sections as well.
The experimental signature of the primitives is zeros modulo $\pi$ of the scattering phase with a negative slope.

Summarizing, the following states are highlighted as zeros of $D$ function:

(i) Stable particles, which we identify with asymptotic states of quantum filed theory 
and register experimentally with the use of detectors. 

(ii) Resonances are observed as peaks in the cross sections. 

(iii) Primitives are observed as zeros modulo $\pi$ of the scattering phase with a negative slope. 

We learn of the existence of resonances and primitives on the basis of observations of stable particles.
The zeros modulo $\pi$ of the scattering phase with a positive slope are identified as the CDD poles 
associated with compound states. Depending on the nature of the interaction, 
compound states are, in turn, associated with bound states, resonances, or primitives.

The concept of a primitive thus leads to a revision of the concept of an elementary particle.
In the broad sense, elementary particles correspond to simple zeros of $ D $ function, 
and only sometimes to both simple zeros of $ D $ function and isolated poles of $ S $-matrix.

In the theory of potential scattering, the negative slope of phase shift indicates repulsion 
between the particles. 
Bound states also tend to reduce the scattering phase. 
We are interested in the nucleon-nucleon scattering, where deuteron is the only bound state. 
In other two-nucleon channels, bound states are absent. In what follows we consider, 
if the opposite is not specified explicitly, $D$ functions without zeros below the threshold.

The $s$-channel exchange of $6q$-primitives was proposed 
to account for the repulsion between nucleons at short distances \cite{SIMO81}. This mechanism
was also found to be sufficient to model the long-distance interaction dominated by the attraction.

The interaction of nucleons is, however,
traditionally described in terms of the $t$-channel exchange of mesons. 
This approach got the greatest advancement in the one-boson exchange (OBE) models.

\begin{figure} [!htb]
\begin{center}
\includegraphics[angle = 0,width=0.382\textwidth]{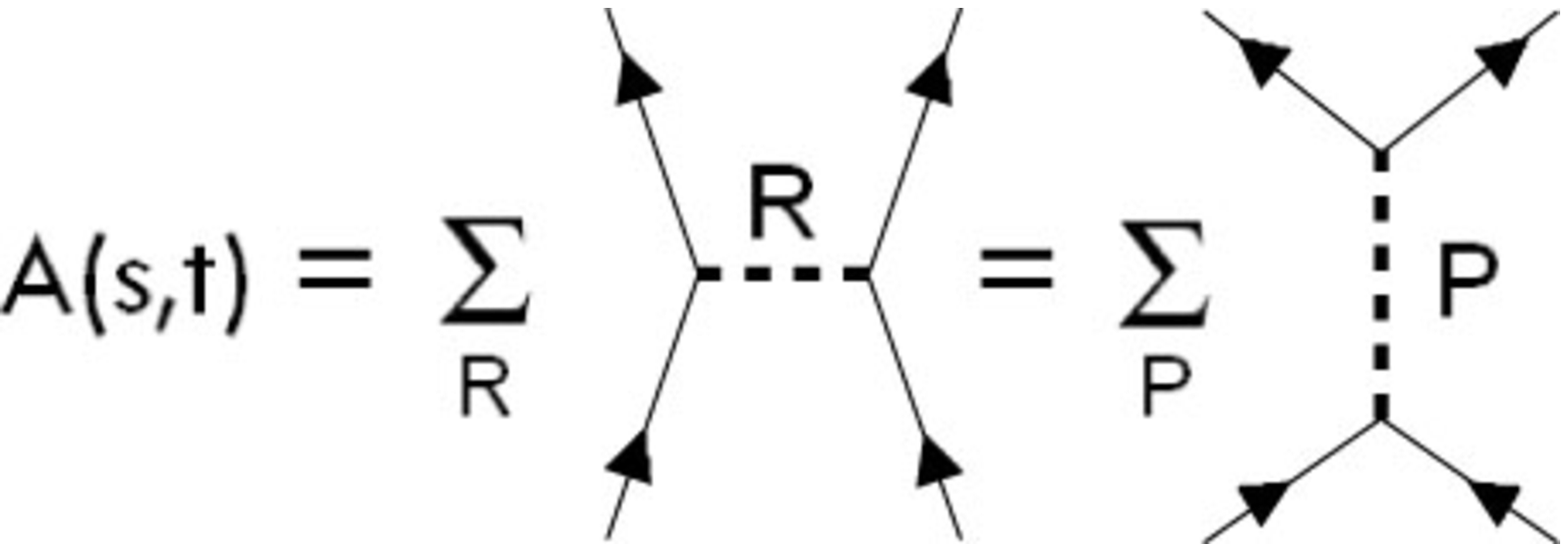}
\caption{
Representation of the scattering amplitude of nucleons as a sum of the $t$-channel meson-exchange diagrams. 
Duality means that the amplitude has an equivalent representation as a sum of primitives ($P$) in the $s$ channel.
$R$ denotes meson resonances. 
}
\label{fig:1}
\end{center}
\end{figure}

Both mechanisms successfully describe a wide range of experimental data. 
This fact is motivating an interest in the possible duality of the $s$-channel exchange of $6q$-primitives 
and the $t$-channel exchange of mesons \cite{MIKH11}.
The duality in the nucleon-nucleon scattering is illustrated in Fig. \ref{fig:1}.


In this paper we investigate conditions under which the duality holds. From a formal point of view, 
it suffices to show that the OBE models can be reformulated in terms of 
an $s$-channel exchange model.

First, however, we discuss in Sect. 2 a hybrid model of Lee \cite{RAJA72}, 
which is not so radically different from the OBE models. 
The $s$-channel exchange of compound states is imbedded in it, but primitives are explicitly absent. 
We examine factors that control the occurrence of 
compound states of the primitive type and provide reducibility to the model \cite{MIKH10}.

The inverse scattering problem for the $s$-channel exchange has many features in common with 
the problem of the reduction of the OBE models. We discuss this problem as well.

In the potential scattering theory and in hybrid Lee models, 
primitives exist only as the peculiarities of scattering phases with no \textit{ab initio} significance.
In Sect. 3, we investigate conditions under which the potential scattering problem can be reformulated 
in terms of an $s$-channel exchange of primitives.
In cases where this is possible, the
primitives become key objects upon the reduction.

\section{Reducibility of hybrid Lee model}

The hybrid Lee model \cite{RAJA72} and the models of Refs. \cite{SIMO81,DYSO57,MIKH10,MIKH11}
are generalizations of the Lee model \cite{TDLE54}. 
We formulate the hybrid Lee model in a language that is convenient for the subsequent reduction.

\subsection{Hybrid Lee model}

The hybrid Lee model is formulated for $n_{c}$ compound states with masses 
$M_{\alpha }$ and $n_{f}$ contact four-fermion vertices. The summation of
the $s$-channel loops gives rise to a $\mathcal{D}$ matrix 
\begin{equation}
\mathcal{D}_{\alpha \beta }(s)=\mathcal{K}_{(\alpha )}\delta _{\alpha \beta
}-\mathcal{P}_{\alpha \beta }(s)  \label{DMAT}
\end{equation}
of the dimension $n=n_{c}+n_{f}$. 
$\mathcal{K}_{(\alpha )}=s-M_{\alpha}^{2}$ are the inverse free propagators of the compound states 
$\alpha =1\ldots n_{c}$ and $\mathcal{K}_{(\alpha )}=C_{\alpha }$ are real constants
entering the vertices of four-fermion interactions $\alpha =n_{c}+1\ldots n$. 
The self-energy operator has the form 
\begin{equation}
\mathcal{P}_{\alpha \beta }(s)=-\frac{1}{\pi }\int_{s_{0}}^{+\infty }\Phi
_{2}(s^{\prime })\frac{\mathcal{F}_{\alpha }(s^{\prime })\mathcal{F}_{\beta
}(s^{\prime })}{s^{\prime }-s}ds^{\prime }.  \label{POLA}
\end{equation}
The integral converges provided $\mathcal{F}_{\alpha }(s)=O(1/s^{\nu })$ at $%
s\rightarrow +\infty $ with $\nu >0$. For $0<\nu <1/2$ 
$\mathcal{P}_{\alpha \beta}(s)=O(1/s^{2\nu })$ at $s\rightarrow -\infty $, and for $1/2<\nu $ 
$\mathcal{P}_{\alpha \beta }(s)=O(1/s)$ at $s\rightarrow -\infty $. The matrix elements 
$\mathcal{D}_{\alpha \beta }(s)$ are analytic functions in the complex 
$s$-plane with the unitary cut $(s_{0},+\infty )$. The discontinuity equals 
\begin{equation}
\Im \mathcal{D}_{\alpha \beta }(s)=-\Im \mathcal{P}_{\alpha \beta }(s)=\Phi
_{2}(s)\mathcal{F}_{\alpha }(s)\mathcal{F}_{\beta }(s).
\label{factor}
\end{equation}
It determines the off-shell decay widths of the compound states $\alpha = 1,\ldots,n_c$:
\[\sqrt{s}\Gamma _{\alpha }(s) = \Im \mathcal{D}_{\alpha \alpha }(s)\geq 0,\] 
and a background for $\alpha = n_c + 1,\ldots,n$.

$\mathcal{P}_{\alpha \beta }(s)$ describes the interaction of  the channels with the continuum. 
The form factors $\mathcal{F}_{\alpha }(s)$ are real functions on the unitary cut.
Compound states correspond to bound states and resonances. 
In the presence of primitives, the correspondence becomes ambiguous.

The scattering amplitude has the form 
\begin{equation}
A(s)=-(\Im \mathcal{D}_{\beta \alpha }(s))\mathcal{D}_{\alpha \beta
}^{-1}(s).  \label{AMPL}
\end{equation}
It is not difficult to see that this amplitude satisfies the unitarity: 
\begin{eqnarray*}
A(s)-A^*(s) &=&2i\Im A(s) \\
&=&-\Phi _{2}(s)\mathcal{F}_{\alpha }(s)\mathcal{D}_{\alpha \beta }^{-1}(s)\mathcal{F}_{\beta }(s)+\Phi _{2}(s)\mathcal{F}_{\alpha }(s)\mathcal{D}_{\alpha \beta }^{*-1}(s)%
\mathcal{F}_{\beta }(s) \\
&=&\Phi _{2}(s)\mathcal{F}_{\alpha }(s)\mathcal{D}_{\alpha \beta }^{*-1}(s)\left(
\mathcal{D}_{\beta \gamma }(s)-\mathcal{D}_{\beta \gamma }^*(s)\right) \mathcal{D}_{\gamma \delta}^{-1}(s)\mathcal{F}_{\delta }(s) \\
&=&2i\Phi _{2}(s)\mathcal{F}_{\alpha }(s)\mathcal{D}_{\alpha \beta }^{*-1}(s)\left(
\Phi _{2}(s)\mathcal{F}_{\beta }(s)\mathcal{F}_{\gamma }(s)\right) \mathcal{D}_{\gamma
\delta }^{-1}(s)\mathcal{F}_{\delta }(s) \\
&=&2i\left| \Phi _{2}(s)\mathcal{F}_{\alpha }(s)\mathcal{D}_{\alpha \beta }^{-1}(s)%
\mathcal{F}_{\beta }(s)\right| ^{2} \\
&=&2i\left| A(s)\right| ^{2},
\end{eqnarray*}
and can therefore be represented as $A(s)=e^{i\delta (s)}\sin
\delta (s)$. The $S$ matrix has the form 
\begin{equation}
S(s) = 1 + 2iA(s) = \mathcal{D}_{\beta \alpha }^*(s)\mathcal{D}_{\alpha \beta }^{-1}(s). 
\label{DSMATP}
\end{equation}
Here and everywhere below, for any function on the real axis of the complex $s$-plane $F(s)$ means $F(s + i0)$, 
$F^*(s)$ means $F(s-i0)$. As analytic function, $F(s+i0)$ is defined on the first sheet of the Riemann surface. 
Equation (\ref{DSMATP}) generalizes Eq.~(\ref{DSMAT}).

Using equation 
\begin{equation*}
\mathcal{D}_{\beta \alpha }^{-1}(s)\det ||\mathcal{D}(s)||=
\frac{1}{(n-1)!}\epsilon _{\alpha \alpha _{2}..\alpha _{n}}
\epsilon _{\beta \beta_{2}...\beta _{n}}
\mathcal{D}_{\alpha _{2}\beta _{2}}(s)\mathcal{D}_{\alpha
_{3}\beta _{3}}(s)...\mathcal{D}_{\alpha _{n}\beta _{n}}(s), 
\end{equation*}
where $\epsilon _{\alpha _{1}\alpha _{2}..\alpha _{n}}$ is the totally
antisymmetric tensor, $\epsilon _{12\ldots n}=+1$, one can write the $N/D$
representation: 
\begin{equation}
A(s)= \frac{\mathcal{N}(s)}{\det ||\mathcal{D}(s)||}. 
\end{equation}
The numerator function has the form 
\begin{equation}
\mathcal{N}(s)=-\frac{1}{(n-1)!}\epsilon _{\alpha _{1}\alpha _{2}..\alpha
_{n}}\epsilon _{\beta _{1}\beta _{2}...\beta _{n}} 
(\Im \mathcal{D}_{\alpha_{1}\beta _{1}}(s))\mathcal{D}_{\alpha _{2}\beta _{2}}(s)...\mathcal{D}%
_{\alpha _{n}\beta _{n}}(s).  \label{NFUN}
\end{equation}
One can verify that, by virtue of Eq.(\ref{factor}), $\Im \mathcal{N}(s)=0$ on the unitary cut, 
and so $\mathcal{N}(s)$ is an analytic function in the
complex $s$-plane except for singular points of the phase space function 
$\Phi _{2}(s)$ and the form factors $\mathcal{F}_{\alpha }(s)$. 
The singularities of $\mathcal{F}_{\alpha }(s)$ originate from the exchange of
particles in the crossing channels.
In Eq.(\ref{NFUN}) $\mathcal{D}_{\alpha \beta }(s)$ can be replaced with $\Re \mathcal{D%
}_{\alpha \beta }(s)$.

The imaginary part of the determinant, 
\begin{equation}
\det ||\mathcal{D}(s)||=\frac{1}{n!}\epsilon _{\alpha _{1}\alpha
_{2}..\alpha _{n}}\epsilon _{\beta _{1}\beta _{2}...\beta _{n}}\mathcal{D}%
_{\alpha _{1}\beta _{1}}(s)\mathcal{D}_{\alpha _{2}\beta _{2}}(s)...\mathcal{%
D}_{\alpha _{n}\beta _{n}}(s),  \label{DETD}
\end{equation}
gives the numerator function on the unitary cut 
\begin{equation}
\Im \det ||\mathcal{D}(s)||=-\mathcal{N}(s).  \label{imde}
\end{equation}

Suppose we have on the real $s$-axis above $s_{0},$ $n_{+}$ simple zeros of $\mathcal{N}(s)$ with
positive $\delta^{\prime}(s)$ and $n_{-}$ simple zeros of $\mathcal{N}(s)$
with negative ones. The zeros of the first and second kinds are denoted
by $s_{+\beta}$ with $\beta = 1,\ldots,n_+$ and $s_{-\beta}$ with $\beta = 1,\ldots,n_-$, respectively. 
Hence we have $\delta(s_{\pm \beta})=0$~$\mod(\pi)$ and $\delta^{\prime}(s_{\pm \beta}) \gtrless 0$.

Levinson's theorem establishes a connection between the asymptotic value of the scattering 
phase and the number of bound states $n_{B}$: 
\begin{equation}
\delta(+\infty) - \delta(s_0) = - \pi n_{B}. 
\label{levi}
\end{equation}
The values of $ n_{+}$ and $ n_{-} $ are related to $n_{B}$. By continuity, 
there are three possibilities:
\begin{equation}
n_{-} - n_{+} - n_{B}= \left\{ 
\begin{array}{rlll}
-1, & \mathrm{if} & \delta^{\prime}(s_0) < 0, &  \\ 
0, & \mathrm{if} & \delta^{\prime}(s_0) \gtrless 0,&  \\ 
1, & \mathrm{if} & \delta^{\prime}(s_0) > 0, &
\end{array}
\right. 
\label{NC1}
\end{equation}
the prime denotes differentiation with respect to the momentum.

\subsection{Simonov-Dyson model}

Noticeable simplifications arise when the form factors coincide up to normalization
\begin{equation}
\mathcal{F}_{\alpha }(s)=g_{\alpha }\mathcal{F}(s).  
\label{FORM}
\end{equation}
In such a case, the determinant of $\mathcal{D}_{\alpha \beta }(s)$ can easily be found. 
Its polynomial part is the product of $\mathcal{K}_{(\alpha )}$, while the dispersion
part is linear in 
\begin{equation}
\mathcal{P}_{\alpha \beta }(s) = g_{\alpha}g_{\beta}\Pi (s).
\label{FACT}
\end{equation}
The result can be written in the form
\begin{equation}
\det ||\mathcal{D}(s)||=\prod_{\alpha }\mathcal{K}_{(\alpha )}\left( 1-\Pi
(s)\Lambda ^{-1}(s)\right),  
\label{DFUN-SD}
\end{equation}
where
\begin{equation}
\Lambda ^{-1}(s)=\sum_{\alpha =1}^{n_{c}}\frac{g_{\alpha }^{2}}{s-M_{\alpha}^{2}}+f.
\label{LAMB-SD}
\end{equation}
The constant $f$ is expressed through the constants $%
C_{\alpha }$ of the four-fermion interaction and $g_{\alpha }$:
\begin{equation}
f=\sum_{\alpha =n_{c}+1}^{n}\frac{g_{\alpha }^{2}}{C_{\alpha }}.
\end{equation}

The $S$ matrix is the ratio of $D$ function on two different sheets of the Riemann surface (see Eq. (2)), 
so multiplication or division of $D$ by a polynomial is a valid
operation. The canonical $D$ function can be taken in the form
\begin{equation}
D(s)=\Lambda (s)-\Pi (s).
\label{DFUN-SD}
\end{equation}

Dispersion integral is determined up to a polynomial. By the replacement $\Pi(s)\rightarrow \Pi (s)+C$ and 
$\Lambda (s)\rightarrow \Lambda (s)+C,$ that
leaves the $D$ function invariant, one can redefine masses of the compound states and shift the $P$-matrix poles.
The masses of the compound states are therefore model dependent.
They are, however, located strictly between the consecutive CDD poles.

From a formal point of view, the distinction of the model \cite{SIMO81} from the model 
\cite{DYSO57} is very subtle. In the latter $\mathcal{F}^2(s) >0$, 
whereas in the former $\mathcal{F}^2(s) \geq 0$. 

In the model of Simonov, the zeros of $\mathcal{F}(s)$ on the unitary cut are thus permitted. 
Their existence is a necessary condition 
for the existence of the primitives. In terms of the $s$-channel exchange, repulsion is generated by the primitives 
and is related to the zeros of $\mathcal{F}(s)$. 

\subsection{Representation $\Im D(s)\geq 0$}

Consider the amplitude (\ref{AMPL}) in the vicinity of $s=s_{+\beta}>s_{0}$. Let $\mathcal{N}%
(s)\varpropto s-s_{+\beta},$ then $\det ||\mathcal{D}(s)||\approx a-i(s-s_{+\beta}).$
Dividing $\mathcal{N}(s)$ and $\det ||\mathcal{D}(s)||$ by $-s+s_{+\beta},$ one
gets 
\begin{equation}
A(s)\approx - \frac{1}{-\frac{a}{s-s_{+\beta}}+i}. 
\label{cdd}
\end{equation} 
The phase shift behaves like $\delta (s)\approx (s-s_{+\beta})/a$~$\mod(\pi)$. 
Using condition $\delta^{\prime }(s_{+\beta})=1/a>0$ we conclude that $a$ is positive.
As compared to the imaginary part, the
residue of the pole term in the denominator of $A(s)$ has the correct sign
for a CDD pole. 

Now, consider $A(s)$ in the vicinity of $s=s_{-\beta}>s_{0}$. Let $\mathcal{N}%
(s)\varpropto -s+s_{-\beta},$ then $\det ||\mathcal{D}(s)||\approx a+i(s-s_{-\beta}).$
Using condition $\delta^{\prime }(s_{-\beta}) < 0$ we conclude that $a$ is positive
and $A(s)\approx -1/(\frac{a}{s-s_{-\beta}}+i)$. The phase shift $\delta
(s)\approx -(s-s_{-\beta})/a$~$\mod(\pi)$ has a negative slope. As compared to the imaginary
part, the residue of the pole term in the denominator of $A(s)$ has the
wrong sign for a CDD pole. One can, however, multiply $\mathcal{N}(s)$ and 
$\det ||\mathcal{D}(s)||$ by $(s-s_{-\beta})/a$ to give 
\begin{eqnarray}
A(s)&\approx&  - \frac{1}{\frac{a}{s-s_{+\beta}}+i} \nonumber \\
&\equiv& - \frac{(s-s_{-\beta})^{2}/a}{s-s_{-\beta}+i(s-s_{-\beta})^{2}/a}. 
\label{prim}
\end{eqnarray}
The amplitude is proportional to the propagator of particle with mass $M_{\beta} = \sqrt{s_{-\beta}}$
and energy dependent width $\sqrt{s}\Gamma (s)\approx $ $(s-s_{-\beta})^{2}/a$. 
The width vanishes on the mass shell. This is signature of primitive.
A quantity $\sim (s-s_{-\beta})^{2}$ in the numerator can be regarded as an expansion 
of the primitive form factor squared near its mass where the form factor vanishes.

The $s_{+\beta}$ and $s_{-\beta}$ zeros should therefore be treated differently. The
canonical $N(s)$ and $D(s)$ functions can be taken to be 
\begin{eqnarray}
N(s) &=& \eta \frac{\pi_{-}(s)}{\pi_{+}(s)}\mathcal{N}(s),  
\label{NCAN}
\\
D(s) &=& \eta \frac{\pi_{-}(s)}{\pi_{+}(s)}\det ||\mathcal{D}(s)||,
\label{DCAN}
\end{eqnarray}
where
\begin{equation}
\pi_{\pm}(s) = \prod\limits_{\beta =1}^{n_{\pm}}(s-s_{\pm \beta }).
\label{rho}
\end{equation}
The analytical properties of $\Im D(s)=-N(s)$ are the same as of $\mathcal{N}(s)$, 
$\Im D(s)$ does not change the sign on the unitary cut. 
Above the threshold $\mathcal{N}(s)$ is real, but not a sign-definite function. The expression
${\pi_{-}(s)} \mathcal{N}(s)/{\pi_{+}(s)}$ is a sign-definite function. Yet it can be both positive and negative. 
We multiplied the right-hand sides of Eqs. (\ref{NCAN}) and (\ref{DCAN}) by $\eta =\pm 1$ to make $N(s)$ non-positive. 
According to the convention, $\Im D(s)=-N(s)\geq 0$.

We thus made the identity transformation, which resulted in the non-negative imaginary part of $D(s)$. 
$\Im D(s)$ can therefore be interpreted in terms of the decay width.

$D(s)$ has $n_{+}$ poles at $s=s_{+\beta }$. The residues of $D(s)$ at 
$s=s_{+\beta }$ are real, since $\Im \det ||\mathcal{D}(s_{+\beta })||$ vanishes. 
$\Re D(s)$ and $\Im D(s)$ have $n_{-}$ first- and second-order zeros at $s=s_{-\beta }$.
In addition, $\Re D(s)$ is zero everywhere, where $\Re \det ||\mathcal{D}(s)||$ vanishes.

We thus arrive at the representation (\ref{DFUN-SD}) of the $D$ function. The dispersion part 
takes the form
\begin{equation}
\Pi (s) = -\frac{s^{\gamma} }{\pi }\int_{s_{0}}^{+\infty }\Phi_{2}(s^{\prime })
\frac{\mathcal{F}^{2}(s^{\prime })}{s^{\prime }-s}\frac{ds^{\prime }}{s^{\prime \gamma} }.  \label{POLAP} \\
\end{equation}
The leading term $\sim s^{n_{c}}$ of $\det ||\mathcal{D}(s)||$ at infinity arises in calculating 
the determinant because of the multiplication of the diagonal elements. 
The imaginary part of $\det ||\mathcal{D}(s)||$ equals 
$O(s^{n_{c} - 1}\Im \mathcal{D}_{\alpha \beta }(s))$ 
and $O(s^{n_{c}}\Im \mathcal{D}_{\alpha \beta }(s))$ at $s\rightarrow +\infty $ for $n_{f} = 0$ and $1 \leq n_{f}$.
Consequently,
we write the dispersion integral with the subtraction. If the coefficient $\gamma$ is chosen as 
$\gamma =n_{c} - 1  + n_{-} - n_{+}$ and
$\gamma =n_{c}  + n_{-} - n_{+}$  for $n_{f}=0$ and $1 \leq n_{f}$, respectively, the convergence of the dispersion
integral in Eq.~(\ref{POLAP}) is the same as in Eq.~(\ref{POLA}).

The form factor $\mathcal{F}(s)$ satisfies the equation 
\begin{equation}
\Phi _{2}(s)\mathcal{F}^{2}(s)\equiv \Im D(s)=- \eta \frac{\pi_{-}(s)}{\pi_{+}(s)} \mathcal{N}(s)\geq 0. 
\end{equation}
$\mathcal{F}(s)$, together with its first derivative, can be taken a continuous function on the unitary cut. 
The form factor defined this way  
becomes an analytic function in the complex $s$-plane, whose singularities are the singularities 
of $\Phi _{2}(s)$ and $\mathcal{N}(s)$. $\mathcal{F}(s)$ has simple zeros at $s=s_{-\beta }$.

$\Lambda (s)$ is a non-dispersive part of $D$ and a rational function. 
Its simple poles are located at $s = s_{+\beta}$.
The residues have negative sign, so the poles are the CDD poles. 

Under certain conditions, which we have to clarify, 
zeros of $\Lambda (s)$ are associated with masses of compound states.
In our case, these include the values of $s_{-\beta }$ and perhaps additional roots of equation $\Lambda(s) = 0$. 
$n­_{*}$ zeros of the second type are also denoted by $s_{-\beta}$, 
the index $\beta$ takes the values $n_{-}+1,\ldots,n_{-} + n_{*}$.
The polynomial part of $\det ||\mathcal{D}(s)||$ has the degree $n_c$, 
so one can expect that $n_* = n_c $.

$\Lambda(s)$ is thus the ratio of two polynomials of the degrees $n_{-} + n_{*}$ and $n_{+}$. 
This function may be provided the representation of (\ref{LAMB-SD}) for
\begin{equation}
n_{-}-n_{+} +n_{*}= \left\{ 
\begin{array}{rlll}
0, & \mathrm{if} & f\neq 0, & \\ 
1, & \mathrm{if} & f=0. & 
\end{array}
\right. 
\label{NC2}
\end{equation}
Equations (\ref{NC1}) and (\ref{NC2}) being combined give strong limit $n_c = n_* \leq 2$.

The structure of $\Lambda (s)$ from Eq.~(\ref{LAMB-SD}) is very simple. Zeros and poles of $\Lambda (s)$ alternate. 
This restriction should be imposed on the non-dispersive part of the $D$ function (\ref{DCAN}). 
If between two consecutive CDD poles only one zero of $\Lambda (s)$ exists, 
the reducibility of the hybrid Lee model to the Simonov-Dyson model is guaranteed.


\subsection{What if there is no reducibility?}

It is not difficult to imagine a situation in which the relations (\ref{NC2}) are not fulfilled.
For example, the phase shift can behave as $\delta(s) = - kb$. 
In potential scattering theory, this phase describes the scattering of a particle by an infinitely 
high potential wall located at a distance $ b $ from the origin.
We leave aside the question of whether it is possible to 
generate such a phase shift
in the Lee model.
In the case of $\delta(s) = - kb$, we have an infinite number of intersections of the levels $- n \pi$ with a negative slope. 
In the vicinity of each such point, the scattering amplitude is given by Eq.~(\ref{prim}). 
We conclude, therefore, that the amplitude there is dominated by the $s$-channel exchange of a primitive 
with mass $2\sqrt{m^2 + (n \pi/b)^2}$, where $m$ is the nucleon mass. 
Until now it was assumed that the number of zeros of the amplitude 
with a negative slope is finite. 
For $\delta(s) = - kb$, however, $n_{-} =  + \infty $ and $ n_ {+} = n_ {*} = 0 $. 
We thus went beyond assumptions of the models discussed.
\footnote{In the model \cite{SIMO81} with infinite number of the zeros, $\delta(s) = - kb$ can be reproduced asymptotically for $s \to +\infty$.}

When the interaction with the continuum is switched off, $\Pi (s) \equiv 0$,
each compound state becomes zero of the $D$ function. In this sense, compound states can be regarded 
as bare particles of the $s$-channel. 

However, there may exist zeros of the $D$ function which do not correspond to compound states. 
These are due solely to the interaction.
In general, only a fraction of zeros of the $D$ function corresponds 
to compound states and hence the $P$-matrix poles. 
Another part disappears when the interaction is switched off. Such primitives match 
neither compound states nor $P$-matrix poles.

In all cases, however, we have the decomposition (\ref{prim}), 
so that there are reasons to highlight contributions of primitives,
regardless of whether the scattering problem admits solution in terms of 
an $s$-channel exchange model.

\subsection{Benjamins-van Dijk model}

Benjamins and van Dijk \cite{BENJ86} used a hybrid model of Lee with $n_{c}=n_{f}=1$ 
to describe the $S$-wave nucleon-nucleon interaction with the aim of constraining the
admixture of the elementary particle component of the deuteron. 
The compound state masses are listed in Table 1. The numerator function 
$\mathcal{N}(s)$ has two zeros on the unitary cut. 
The first low-energy one with $\delta^{\prime }(s)<0$ is of the primitive type. 
The second high-energy one with $\delta^{\prime }(s)>0$ corresponds to the CDD pole. 
We thus have $n_{-} = n_{+} = 1$. 
The zeros occur at $T_{lab}=354$
MeV and at about $700$ MeV in the $^{3}S_{1}$ channel, and at $T_{lab}=265$
MeV and at about $1$ GeV in the $^{1}S_{0}$ channel. This case 
corresponds to $\gamma =0$ and thereby does not require subtraction.

Upon the reduction, the lower zeros of $\mathcal{N}(s)$ determine the masses of primitives, 
while the higher zeros determine the CDD poles. These zeros lead to the existence of 
two compound states in each of the channels. 
Their masses are determined from the equation $\Lambda (s)=0$. 
The second mass $M_{2}$ is, however, very large and can not be determined reliably.
The physical states correspond to solutions of Eq.~(\ref{DZERO}).
They are a primitive and a high-mass resonance. In Table \ref{tab:table1} we
compare parameters of the model \cite{BENJ86} before and after the reduction
with parameters of the model \cite{MIKH10}. 
In the original model, 
$\det ||D(s)||$ does not have CDD poles, while the equation $\det ||D(s)||=0$ does not have
roots on the unitary cut. The CDD poles and the primitives appear upon the
reduction. 


\begin{table} [!htb]
\renewcommand{\arraystretch}{1.0}
\caption{
Parameters of the hybrid Lee model \cite{BENJ86}, the reduced hybrid Lee (RHL) model, and the model \cite{MIKH10} in the $^{3}S_{1}$ and 
$^{1}S_{0}$ channels of the nucleon-nucleon scattering. $M^{[0]}_{i}$ and $M_{i}$ are masses (in MeV) of the compound states and the physical states, respectively.
$M$ is position of the CDD pole (in MeV). 
}
\centering
\label{tab:table1}
\begin{tabular}{|c|c|c|c|c|c|}
\hline \hline
$^3S_1$    & \multicolumn{2}{|c|}{compound states}      & \multicolumn{2}{|c|}{physical states}        & {CDD pole}   \\ \hline 
Model      & $M^{[0]}_{1}$   & $M^{[0]}_{2}$        & $M_{1}$           & $M_{2}$              & $M$          \\ \hline 
\cite{BENJ86}         & 2228              &                        & high                &                        &              \\ 
RHL        & 2167              & $\sim 10^7$       & 2047                & high                   & 2204         \\ 
\cite{MIKH10}         & 2047              &                        & 2047                &                        & 3203         \\ \hline\hline 
$^1S_0$    & \multicolumn{2}{|c|}{compound states}      & \multicolumn{2}{|c|}{physical states}        & {CDD pole}   \\ \hline 
Model      & $M^{[0]}_{1}$   & $M^{[0]}_{2}$        & $M_{1}$           & $M_{2}$              & $M$          \\ \hline 
\cite{BENJ86}         & 2328              &                        & high                &                        &              \\ 
RHL        & 2310              & $\sim 10^5$       & 2006                & high                   & 2321         \\ 
\cite{MIKH10}         & 2006              &                        & 2006                &                        & 2916         \\ 
\hline\hline  
\end{tabular}
\end{table}


The hybrid Lee model in the version of Benjamins and van Dijk
can therefore be reduced to and interpreted physically in terms of the Simonov-Dyson model.

\section{One-boson exchange and inverse scattering problem for $s$-channel}

The specific properties of the Lee model still were used very little. 
Any scattering amplitude can be written as the $N / D$ ratio. 

In our reduction scheme, zeros of $\mathcal{N}(s)$ on the unitary cut play special role. 
They define polynomials $\pi_{\pm}(s)$ according to Eq.~(\ref{rho}). By multiplying 
the initial $N$ and $D$ functions 
with a rational function $\pi_{-}(s)/\pi_{+}(s)$, 
we obtain $D(s)$ with a non-negative imaginary part, CDD poles, and zeros 
in a neighborhood of which $D(s)$ looks like denominator of the Breit-Wigner formula 
with energy-dependent width (see Eq.~(\ref{prim})). 
$\Im D(s) \geq 0$ can be interpreted as the probability. 

Reduction scheme used in Sect. 2.3 is general enough to be applied to the OBE models, 
or even to redefine the phenomenological $ N $ and $ D $ of functions 
constructed on the basis of the experimental phase shifts.

Let the scattering phase be known experimentally in the energy range $(s_{0},+\infty)$. 
We construct the Jost function
\begin{equation}
D_{J}(s) = \exp ( -\frac{s}{\pi}\int_{s_{0}}^{+\infty} \frac{ \delta(s^{\prime})}{s^{\prime}(s^{\prime} - s)}ds^{\prime} ).
\end{equation}
It is assumed that the dispersion integral converges well enough 
so that asymptotically $D_J(s) = O (1)$. 
Analyzing the imaginary part, we isolate its zeros $s_{+\beta}$ and $s_{-\beta}$ 
with positive and negative slopes of the phase shift. 
These zeros are the zeros of the scattering amplitude. After that, we redefine $D_{J}(s)$ 
as in Eq.~(\ref{DCAN}) and introduce 
\begin{equation}
D(s) = \eta \frac{\pi_{-}(s)}{\pi_{+}(s)} D_{J}(s).
\label{rhoDJ}
\end{equation}
As a result, $\Im D(s)$ is non-negative. 
The difference between $D(s)$ and its dispersive part is a rational function $\Lambda(s)$ 
whose poles are the CDD poles. 
The zeros of $\Lambda(s)$ could determine the masses of compound states associated with resonances and primitives 
(as well as bound states and virtual states not discussed here).

It is known that any scattering data set can be associated with a smooth potential whose physical origin lies in the $t$-channel exchange.

A sufficient condition for the existence of an $s$-channel exchange model, in which the scattering phase $\delta(s)$ is reproducible, 
is the representation of $\Lambda(s)$ in the form of the function (\ref{LAMB-SD}). 
This is possible if zeros and poles of $\Lambda(s)$ alternate.
Under these conditions the inverse scattering problem can successfully be solved 
in the sense of finding the physically meaningful parameterization of the $s$-channel exchange interaction. 

An example of such a solution is the reduction of hybrid Lee model discussed in Sect. 2.5 
provided we consider the scattering phase as empirical quantity. 
The $D$ function constructed there has the correct analytical properties and, up to normalization, coincides therefore 
with the function (\ref{rhoDJ}).

In Ref. \cite{MIKH10} the empirical phases of nucleon-nucleon scattering in the $^3 S_1$ and $^1 S_0$ channels
are described using the model \cite{SIMO81} 
after imbedding correct analytic properties. 
We further describe the nucleon-nucleon scattering in the 
$^3 P_1$, $^1 P_1$ and $^3 P_0$ channels \cite{NADE11}. In the cases considered, 
the correct analytical properties are satisfied, 
so that the constructed $D$ functions essentially coincide with the function (\ref{rhoDJ}),
up to normalizations,
neglecting inelasticity, and the incomplete knowledge of the experimental phase shifts. 
In Refs. \cite{MIKH10,NADE11}, the inverse scattering problem 
is not posed literally because of involvement of \textit{a priori} form factor 
entering the vertices of compound states and
the contact four-fermion vertices.
Masses of compound states and their couplings with the continuum, as well as the four-fermion couplings, 
are however determined completely in the spirit of 
the inverse scattering procedure for an $s$-channel exchange interaction.

\section{Conclusion}

The primitives are simple zeros of $D$ function on the unitary cut in the normalization where the imaginary part of $D$ is non-negative 
and the only poles of $D$ are the CDD poles.
In the vicinity of the primitive the scattering amplitude is described by the Breit-Wigner formula. 
Specificity of primitives is the disappearance of their width on the mass shell.

In this paper, we found conditions under which OBE models with the $t$-channel meson exchange 
are dual to the Simonov-Dyson model where the nucleon-nucleon interaction is provided by the $s$-channel exchange of $6q$-primitives.
The analysis is based on the $N/D$ representation of the scattering amplitude and the possibility 
to redefine the $ N $ and $ D $ functions so as to meet the requirement of the non-negativity of $\Im D(s)$. 
The proposed technique with no significant changes applies to the inverse scattering problem of the $s$-channel exchange
in the general case. 

We showed that the duality takes place provided the non-dispersive part
of $D$ function has alternating zeros and poles.
This condition also is sufficient for solving the inverse scattering problem for the $s$-channel exchange. 

There are good phenomenological reasons to believe that the nucleon-nucleon interaction 
is of the type of interactions for which the duality holds at least approximately.

The arguments can be much detailed when considering reduction of the Benjamins and van Dijk version of the
hybrid Lee model. This model, as well as the Simonov-Dyson model, 
provides reasonable description of the elastic nucleon-nucleon scattering data. Although sets of the compound states 
are different, the models were found to be equivalent.

A solution of the inverse scattering problem for the $s$-channel exchange mechanism 
was presented for a limited set of systems. 
The problem of constructing an $s$-channel exchange model, 
in which the inverse scattering problem admits solution for any scattering data set, deserves additional study.

\begin{acknowledgement}
M.I.K. was supported by Grant \# 4568.2008.2 of Leading Scientific Schools of Russian Federation. 
\end{acknowledgement}

\end{document}